\documentclass[a4paper,twocolumn]{article}
\usepackage{latexsym}
\usepackage{graphics}

\begin{document}

\title{Cellular Automata: Wolfram's Metaphors for Complex Systems}

\author{Pratip Bhattacharyya\thanks{pratip@cmp.saha.ernet.in}\\
 \small Theoretical Condensed Matter Physics Division,\\
 \small Saha Institute of Nuclear Physics,\\
 \small 1/AF Bidhannagar, Kolkata 700 064, India}

\date{April 27, 2004}

\maketitle

\indent In the late 1940's while he was trying to construct a model for
self-reproduction, a machine equivalent to biological systems, von Neumann
invented a class of discrete mathematical systems called cellular
automata~\cite{vonNeumann1966}.
Almost thirty years after the invention cellular automata underwent a
radical reformation when, in the early 1980's, Wolfram considered them
as general mathematical representations of complex systems in
nature~\cite{Wolfram1983, Wolfram1984a}.
His investigations on cellular automata led him to the conviction that
the laws for complex systems cannot be formulated as conventional
mathematical equations; he proposed that the evolution of these systems
can be correctly described only in the form of algorithms, the kind that
are used in computer programs. It was the beginning of a new branch of
science which Wolfram originally called the science of complexity.
This new branch of science is based on the notion of
computation~\cite{Wolfram1984b}. According to Wolfram the evolution of
any system, natural or artificial, can be viewed as a computation for
which the initial state of the system is the input and the state that
emerges after a given interval of time is the output. Cellular automata
provided the ground for the discovery and the development of this new
science.

\indent Cellular automata are discrete dynamical systems defined on
discrete space -- an array of finite geometric cells, e.g, in the form
of a lattice -- and evolve in discrete time, i.e., time that changes in
finite steps. The state of a cellular automaton is the set of the
simultaneous states of its component cells. The function of each cell
is defined in terms of a discrete dynamical variable that has a finite
set of values and each value in this set denotes a distinct state of
the cell. A cellular automaton evolves by updating the value of the
dynamical variable simultaneously at all the cells forming the discrete
space; this parallel updating is the primary feature of cellular automaton
dynamics. The updating occurs at each step in discrete time. All updates
follow a simple deterministic rule. The rule states, in terms of an
algorithm, how the state of each cell (determined by the dynamical variable)
changes after each time-step, influenced by itself and the states of the
other cells in a well defined neighborhood. In the simplest cases, called
elementary cellular automata, a two-valued variable $x \in \{ 0, 1\}$
evolves simultaneously at all the cells $i$ of a one-dimensional lattice
by means of an updating rule $F$ that defines the interaction of each cell
with its two nearest neighbors:

\begin{equation}
x_i^{(t+1)} = F \left [ x_{i-1}^{(t)}, x_i^{(t)}, x_{i+1}^{(t)} \right ].
\label{eq:elementary-CA}
\end{equation}

\begin{figure}[ht]
\resizebox{!}{!}{\rotatebox{0}{\includegraphics{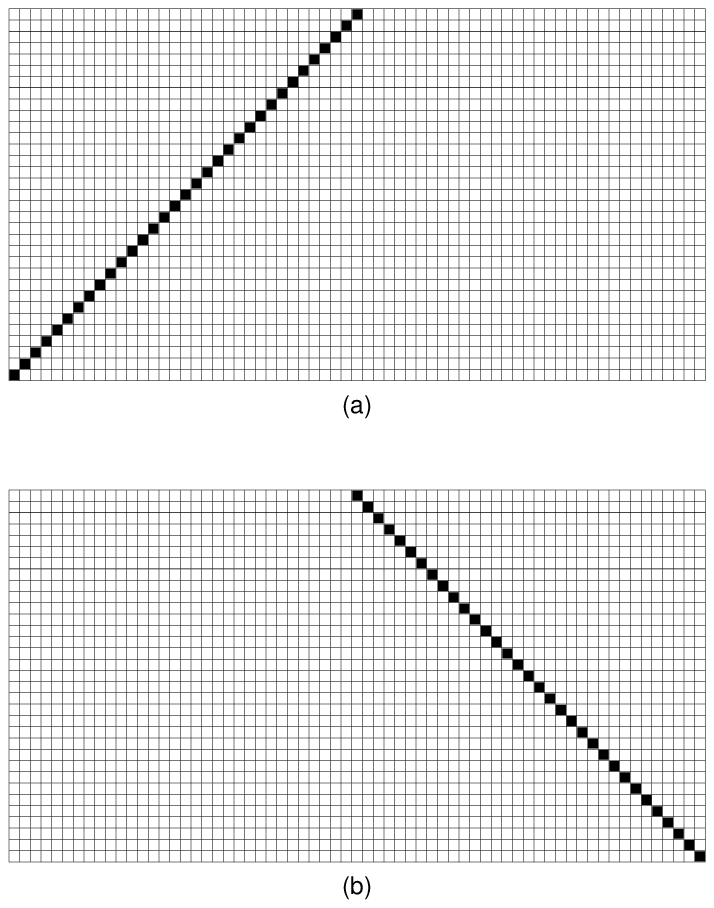}}}
\caption{\small The evolution of two elementary cellular automata with
reversible updating rules: (a) the left shift automaton: $x_i^{(t+1)}
=x_{i+1}^{(t)}$, and (b) the right shift automaton: $x_i^{(t+1)}
=x_{i-1}^{(t)}$. In Wolfram's nomenclature scheme~\cite{Wolfram1983}
these are called rule $170$ and rule $240$ respectively. Each diagram shows
an array of 65 cells evolving for $32$ time-steps. Time increases in the
downward direction. A white square denotes a cell in state $0$ while
a black square denotes a cell in state $1$. In both cases the initial
state of the cellular automaton contains a single cell in state $1$
whereas the rest of the cells are in state $0$. This simple structure
of the initial state is maintained throughout the evolution. If the final
state in each case is considered as the initial state by inverting the
diagrams (equivalent to reversing the direction of time) the evolution of
each automaton is retraced when the respective updating rules are applied;
this happens because the shift automata are reversible.}
\label{fig1}
\end{figure}

\begin{figure}[ht]
\resizebox{!}{!}{\rotatebox{0}{\includegraphics{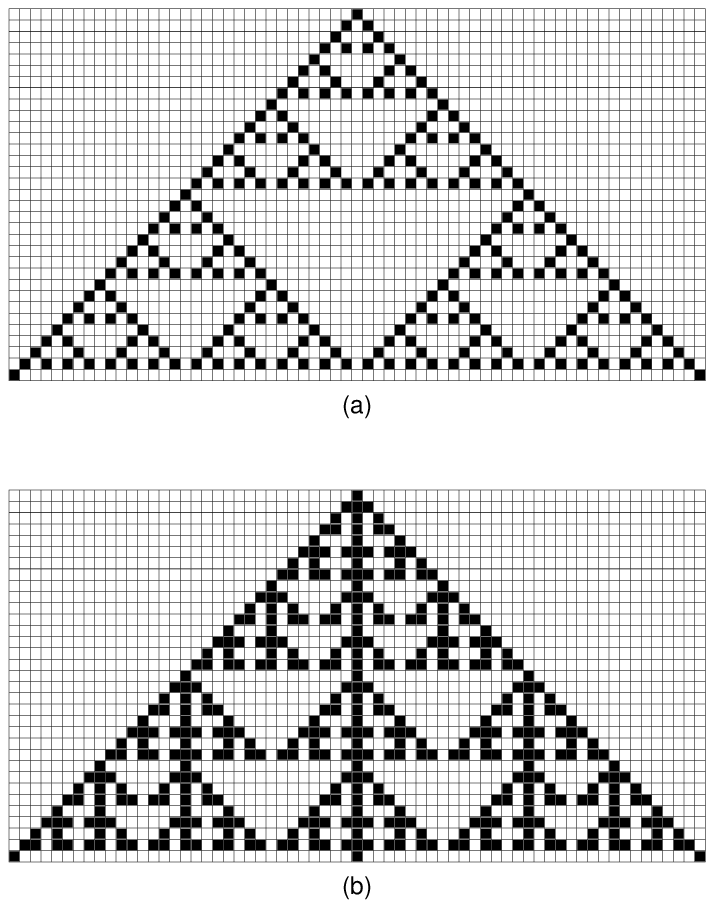}}}
\caption{\small The evolution of two elementary cellular automata with
irreversible updating rules. In Wolfram's nomenclature scheme these
are called (a) rule $90$: $x_i^{(t+1)}=\left [ x_{i-1}^{(t)}
+ x_{i+1}^{(t)} \right ] \bmod 2$, and (b) rule $150$: $x_i^{(t+1)}
=\left [ x_{i-1}^{(t)} + x_i^{(t)} + x_{i+1}^{(t)} \right ] \bmod 2$.
As in Figure 1, white and black squares denote cells in the states $0$
and $1$ respectively and the evolution of an array of $65$ cells is shown
for $32$ time-steps from an initial state which contains a single cell
in state $1$. Time increases downwards. It is clear that the simplicity
of the initial state is destroyed as the automata evolve and an ordered
structure emerges in each case. These cellular automata fail to retrace
their evolution if the direction of time is reversed, which proves that
they are irreversible.}
\label{fig2}
\end{figure}

\noindent Since each cell has two possible states there are $2^3$
different states of the three cell neighbourhood $\left \{ x_{i-1},
x_i, x_{i+1} \right \}$ and each of them maps to a new state of the
central cell $i$ for which there are the same two possibilities as
there are for all other cells. Consequently there are $2^{2^3}=256$
different updating rules $F$ for elementary cellular automata.
Evolution of two different kinds, reversible and irreversible,
are shown in Figs.~\ref{fig1} and~\ref{fig2}. Ref.~\cite{Wolfram1984c}
contains an overview of cellular automata.

\indent The construction of cellular automata was motivated by the
fact that natural systems, both physical and biological, are made of
a large number of elementary units. Each unit has a simple structure
and performs a simple function. However, when these are connected together
by local interactions, the resulting assembly (that forms a natural system)
often produces extremely complex behavior. Besides, natural systems are
inherently dissipative. Consequently they evolve in a manner that is
irreversible and self-organizing, i.e., ordered structures are generated
spontaneously from disordered initial forms. Cellular automata are
mathematical systems that possess similar features. Like natural systems
they are comprised of many identical units, each very simple, that evolve
simultaneously by local interactions into complex ordered states.
The unit of all cellular automata is a geometric cell in discrete space
with a dynamical variable describing its state. Most of the updating rules
are irreversible and generate self-organized states. Wolfram began his
research on cellular automata with the aim of discovering the laws of
self-organization. According to the second law of thermodynamics
an isolated system spontaneously evolves to a state of maximum entropy and
hence, of maximum disorder. The proof of this statement assumes that the
microscopic evolution rule (i.e., the updating rule for each unit of the
system) is reversible. Instead, if the microscopic evolution rule is
irreversible, this particular restriction due to the second law of
thermodynamics no longer exists and Wolfram showed that a system may
evolve from a disordered state to a more ordered one. This is the origin
of self-organization in most cellular automata. At each time-step of
evolution the state of a cellular automaton has a unique successor, since
the updating rule is deterministic. If the rule is reversible the
predecessor of each state is also unique and the set of all allowed states
of the cellular automaton remains constant under its evolution. However,
if the updating rule is irreversible, several distinct states may evolve
to one particular state which means that the predecessor of a state is not
necessarily unique; unless the state in each time-step of evolution is
memorized, the cellular automaton has no way of retracing its history when
the direction of time is reversed. Therefore the set of allowed states of
the cellular automaton contracts as it evolves and the limiting  set of
ordered states that ultimately remains is only a small subset of all
possible initial states. This process of selecting a specific subset of
all possible states forms the mechanism of self-organization. In cellular
automata, the information on the specific subset to be selected is encoded
in the updating rules. Though the updating rules are simple it appears that
the outcomes of the evolution of most cellular automata are impossible
to predict; this, according to Wolfram, is the mark of a complex system.
Wolfram thus adopted cellular automata as the appropriate mathematical
representations of the complex systems occurring in nature. However
Wolfram's definition of a complex system is only qualitative: a system
that is not obviously simple; it still lacks a definition in quantitative
terms.

\indent The evolution of cellular automata are found to be equivalent to
computations, i.e., cellular automata perform like digital computers.
Besides self-organization, this is the other important feature of cellular
automata. In general, each cellular automaton can perform a specific
computation when provided with a specific form of the initial state.
For example, the elementary cellular automaton that follows rule $132$
(in Wolfram's nomenclature scheme~\cite{Wolfram1983}) can effectively
compute the remainder after dividing a natural number $n$ by $2$ if it is
assigned the initial state that contains a block of $n$ consecutive cells
in state $1$ and all other cells on both sides of the block are in state $0$.
The updating rule is expressed as $x_i^{(t+1)}=\left [ x_{i-1}^{(t)}
 x_{i+1}^{(t)} + \left ( 1+x_{i-1}^{(t)} \right ) \left ( 1+x_{i+1}^{(t)}
\right ) \right ] x_i^{(t)} \bmod 2$.
The outcome of the evolution of this cellular automaton tells whether a given
natural number $n$ is even or odd. If $n$ is even, the cellular automaton
evolves to a state where all the cells are in state $0$; if $n$ is odd,
it evolves to a state that contains a single cell in state $1$.
Some cellular automata are known to be capable of universal computation,
i.e., they can perform any possible computation with appropriate initial
states. One of the earliest known examples is the two dimensional cellular
automaton \lq Life\rq~invented by Conway~\cite{Conway1970, Gardner1970}.
The simplest of all those that have been proved to be universal is the
cellular automaton that follows elementary rule $110$: $x_i^{(t+1)} =
\left [ \left ( 1 + x_{i-1}^{(t)} \right ) x_i^{(t)} x_{i+1}^{(t)}
 + x_i^{(t)} + x_{i+1}^{(t)} \right ] \bmod 2$.
The proof is indirect: the elementary cellular automaton with rule $110$
was shown to emulate any given cyclic tag system and it was possible
to construct a cyclic tag system that emulates any given Turing machine;
since some Turing machines are known to be universal computers, it
establishes that rule $110$ is capable of universal computation.
The computational capability of cellular automata led Wolfram to
the view that all processes in nature are programs composed of simple
algorithms in the form of cellular automata.

\indent Wolfram's research on cellular automata for almost two decades has
been recorded in his book \lq A New Kind of Science\rq~\cite{Wolfram2002}.
The book is an outstanding collection of computer experiments
and each set of experiments culminates in a remarkable discovery
or proposition, two of which must be mentioned. While studying the
evolution of reversible cellular automata from various initial states,
Wolfram discovered that there are certain reversible cellular automata
that, contrary to the existing belief, do not obey the second law of
thermodynamics. Some of these reversible automata never evolve to
disordered states from ordered ones; there are others that are found to
self-organize from disordered initial states to configurations
with ordered structures that are reminiscent of the outcomes of
irreversible evolution. Though Wolfram has studied a vast number of
cellular automata, the laws of self-organization have not been found.
However, the results of these computer experiments led him to propose
\lq the principle of computational equivalence\rq. In a general way,
the principle states that \lq almost all processes that are not
obviously simple can be viewed as computations of equivalent
sophistication\rq~\cite{Wolfram2002}. The principle makes a remarkable
assertion that there is just one level of computational sophistication.
Though it is still in the form of a hypothesis, Wolfram believes that
this principle is a new law of nature.

\begin{small}
\textbf{Acknowledgments} I thank Arnab Chatterjee and Sanjay Gupta for
their comments on this article.
\end{small}

\end{document}